\documentclass[12pt,preprint]{aastex}
\begin{document}

\title{A Three-Decade Outburst of the LMC Luminous Blue Variable R127
Draws to a Close}

\author{Nolan~R.~Walborn,\altaffilmark{1}
        Otmar~Stahl,\altaffilmark{2}
        Roberto~C.~Gamen,\altaffilmark{3}
        Thomas~Szeifert,\altaffilmark{4}
        Nidia~I.~Morrell,\altaffilmark{5}
        Nathan~Smith,\altaffilmark{6}
        Ian~D.~Howarth,\altaffilmark{7}
        Roberta~M.~Humphreys,\altaffilmark{8}
        Howard~E.~Bond,\altaffilmark{1} and
        Daniel~J.~Lennon\altaffilmark{1}\\
\vspace{.75in}}        

\altaffiltext{1}{Space Telescope Science Institute, 3700 San Martin Drive,
Baltimore, MD 21218, USA.  STScI is operated by AURA, Inc., under NASA 
contract NAS 5-26555.  
\email{walborn@stsci.edu, bond@stsci.edu, lennon@stsci.edu}}

\altaffiltext{2}{ZAH, Landessternwarte K\"onigstuhl, 69117 Heidelberg, Germany.
\email{O.Stahl@lsw.uni-heidelberg.de}}

\altaffiltext{3}{Complejo Astron\'omico El Leoncito, Avda. Espa\~na 1412
Sur, J5402DSP, San Juan, Argentina.  Member of Carrera del Investigador,
CONICET, Argentina.
\email{rgamen@casleo.gov.ar}}

\altaffiltext{4}{European Southern Observatory, Alonso de C\'ordova 3107, 
Santiago, Chile.
\email{tszeifer@eso.org}}

\altaffiltext{5}{Las Campanas Observatory, The Carnegie Observatories,
Casilla 601, La Serena, Chile.
\email{nmorrell@lco.cl}}

\altaffiltext{6}{Astronomy Department, University of California,
601 Campbell Hall, Berkeley, CA 94720, USA.
\email{nathans@astro.berkeley.edu}}

\altaffiltext{7}{Department of Physics and Astronomy, University College 
London, Gower Street, London WC1E 6BT, UK.
\email{idh@star.ucl.ac.uk}}

\altaffiltext{8}{Astronomy Department, 116 Church Street, S.E., 
University of Minnesota, Minneapolis, MN 55455, USA.
\email{roberta@isis.spa.umn.edu}}

\begin{abstract}
The paradigmatic Luminous Blue Variable R127 in the Large Magellanic Cloud 
has been found in the intermediate, peculiar early-B state, and substantially 
fainter in visual light, signaling the final decline from its major
outburst that began between 1978 and 1980.  This transformation was detected 
in 2008 January, but archival data show that it began between early 2005 and
early 2007.  In fact, significant changes from the maximum, peculiar A-type
spectrum, which was maintained from 1986 through 1998, had already begun
the following year, coinciding with a steep drop in visual light.  We
show detailed correspondences between the spectrum and light, in which
the decline mimics the rise.  Moreover, these trends are not monotonic
but are characterized by multiple spikes and dips, which may provide 
constraints on the unknown outburst mechanism.  Intensive photometric and
spectroscopic monitoring of R127 should now resume, to follow the decline
presumably back to the quiescent Ofpe/WN9 state, in order to fully document 
the remainder of this unique observational opportunity.   
\end{abstract}


\section{Introduction}

The Luminous Blue Variable (LBV) phase is believed to represent a critical
transition in the late evolution of all stars with initial masses greater 
than about 50~M$_{\sun}$, during which they lose sufficiently large amounts 
of mass in recurrent explosive events, to subsequently become classical 
WR stars.  Although the physical mechanism producing these outbursts 
remains to be definitively identified, it is likely related to the 
Eddington Limit (Humphreys \& Davidson 1994, Smith \& Owocki 2006), and
there is ample phenomenological evidence to support the opening statement
above.  A fundamental observational characteristic of massive stellar
populations is that the most luminous blue stars have no red counterparts.  
The LBVs lie just to the left of the inclined Humphreys-Davidson 
(HD; 1979, 1984, 1994; Humphreys 1983) Limit in the HR Diagram, beyond 
which no stars are observed; that is, they appear to define the locus of 
an instability that prevents further redward evolution.  During their 
outbursts, LBVs expand and develop denser, optically thick winds
that reprocess the photospheric radiation to lower effective
temperatures, such that they exhibit horizontal excursions across the HD
Limit, with increasingly cooler spectra and brighter visual magnitudes while
maintaining constant bolometric luminosities.  Following an outburst, they
invert these displacements and return to the hotter, minimum spectrum with
fainter visual magnitude.  Most LBVs and numerous related objects now
interpreted as quiescent LBVs have nitrogen-rich circumstellar nebulae,
produced by prior outbursts thousands of years earlier (Walborn 1982;
Stahl \& Wolf 1986a; Stahl 1987; Nota et~al.\ 1992, 1995, 1996; Clampin et~al.\ 
1993; Davidson et~al.\ 1993; Smith, Crowther, \& Prinja 1994; Smith et~al.\ 
1997, 1998; Pasquali, Nota, \& Clampin 1999; Lamers et~al.\ 2001).

$\eta$~Carinae is the most famous and extreme LBV (Davidson \& Humphreys 
1997), but it is located in a giant H~II region associated with some of the 
most massive stars known, well above 100~M$_{\sun}$ (Walborn et~al.\ 2002).  
Given the uncertainty about the LBV mechanism, it is likewise unclear whether 
or not $\eta$~Car represents a more spectacular version of typical LBVs, 
which are located in more evolved regions and have initial masses between 
50 and 100~M$_{\sun}$.  Well observed examples of the latter are AG~Carinae 
(Wolf \& Stahl 1982, Stahl 1986, Hutsem\'ekers \& Kohoutek 1988, Walborn \& 
Fitzpatrick 2000, Stahl et~al.\ 2001); and R71 (Wolf, Appenzeller, \& Stahl 
1981) and S~Doradus (Massey 2000, and references therein) in the Large 
Magellanic Cloud (LMC).  The known incidence of these objects in other
Local Group galaxies has been greatly extended by Massey et~al.\ (2007).

Radcliffe (R)127 (Feast, Thackeray, \& Wesselink 1960) =
HDE~269858\footnote{Feast et~al.\ denoted it as HDE~269858f for
``following'' because of the companion 3$\arcsec$ (north)west, now
measured at $y=12.12$ and classified as B0~Ia (N weak) by Heydari-Malayeri 
et~al.\ (2003).  It is important to note that the HDE-based photographic 
magnitude of 11.4 given by Feast et~al.\ must refer to the combined light
of the pair, whereas their description of it as a ``double'' likely implies
comparable magnitudes, consistently with the foregoing values.  Most or
all of the photometry discussed in this paper includes both stars;
accordingly, all values presented have been corrected for the companion,
which is significant at the fainter LBV magnitudes.} = Sanduleak 
(1970, Sk)~$-69^{\circ}$~220 is arguably the most important typical LBV.
In fact, taking into account this report, it is not an overstatement 
that it is comparable to $\eta$~Car and even SN~1987A, with regard to
what it may teach us about late massive stellar evolution.  Its highly 
peculiar spectrum was classified as O~Iafpe by Walborn (1977; observation
date 1975 Dec), later reformulated to Ofpe/WN9 by Walborn (1982; echelle 
observation 1978 Nov).  A number of similar objects were
found in the LMC (Bohannan \& Walborn 1989), which have subsequently been
reclassified as WN9-11 types by Crowther \& Smith (1997).  In addition to 
the normal Of selective emission lines of He~II $\lambda$4686 and N~III
$\lambda\lambda$4634-4640-4642, these spectra display other emission
features and P~Cygni profiles including Balmer and He~I lines, together with
very narrow N~III and Si~IV absorptions indicative of unusually extended
atmospheres.  In 1982, R127 was discovered to have entered a classical LBV 
outburst (Stahl et~al.\ 1983, Walborn 1983), thus establishing the Ofpe/WN9 
category as a quiescent LBV state.  Several other members of the category 
were subsequently discovered to possess LBV circumstellar nebulae (references 
above), supporting the foregoing generalization.  R127 is located in a compact 
subgroup of a small, evolved cluster (NGC~2055) containing several mid-O 
through early-B supergiants and devoid of nebulosity (Walborn et~al.\ 1991; 
Heydari-Malayeri, Meynadier, \& Walborn 2003), south of 30~Doradus and not far 
from SN~1987A.  This paper reports that R127 has likely now begun its final 
decline from the current outburst; it is thus the largest and most extensive 
LBV event that will have been covered from start to finish with modern 
instrumentation. 

\section{Observations}

The discovery of R127's decline began with inquiries by NRW of several 
specialists at and immediately following IAU Symposium 250 in 2007 December, 
regarding the current status of its monitoring.  When no such program could 
be identified, check observations were requested of and performed by RCG at 
the CASLEO Observatory in Argentina, and by NIM at the Las Campanas 
Observatory (LCO) in Chile.  These data immediately revealed that R127 was 
then in an intermediate, peculiar B-type state with strong He~I P~Cyg 
profiles, as well as N~II, Si~III, and Si~IV features (Walborn et~al.\ 2008), 
completely different from the maximum, peculiar A-type spectrum dominated 
by Fe~II and Ti~II, in which it had been in 2002 when intensive
monitoring by OS and TS ceased.  Direct images from the LCO 1~m Swope
telescope (Fig.~1) requested by NIM and measured by RCG showed a significant 
decline in visual light with respect to the previous AAVSO data.  
Subsequent archival searches and a casual comment by NRW to NS yielded, 
respectively, the critical spectroscopic observations of 2007 February from 
the European Southern Observatory (PI B.~Davies), which shows that R127 was 
already in the early-B state then; and of the unique transitional spectrum 
in 2005 February from LCO.  All of the high-resolution spectroscopic 
observations discussed here are specified in Table~1.  Key segments of 
them are reproduced in Figures~2--4.  A summary lightcurve is presented in
Figure~5, extending from 1982 through early 2008.  The 2002 and earlier
data are representative of many more obtained by OS/TS, whereas the
2005--2008 are all the high-resolution, blue-violet observations of which
we are aware.  Here we shall describe the major spectral transformations 
with reference to the lightcurve.  

\section{Discussion}

The R127 outburst through 1986 was extensively documented by Stahl et~al.\ 
(1983), Stahl \& Wolf (1986b), and Wolf et~al.\ (1988).  As they discuss, the
spectroscopic observation by Walborn (1982) and subsequent photometry limit
the onset of the outburst to the interval 1978--1980.  Most of the
spectroscopic data in the early papers were photographic; only the digital
data are now available in electronic form and are shown here; thus it is
essential to refer to the figures in those papers for the complete early
record, and also for detailed line identifications in the A-type spectra.

The early outburst photometry twice appeared to show declines from maxima, 
in 1982 and 1984, but these are now seen to be only inflections during the 
rise to the true maximum (Fig.~5), when R127 became the brightest star in 
the LMC.  The first outburst spectroscopic observation, in 1982 January, 
showed a peculiar early B-type spectrum, similar but not identical to the 
current state (Stahl et~al.\ 1983).  Subsequently (Stahl \& Wolf 1986b), 
the spectrum changed to $\sim$B7 (based upon near equality of He~I 
$\lambda$4471 and Mg~II $\lambda$4481; Walborn \& Fitzpatrick 1990) in 
1983 February during the decline from the 1982 lightcurve peak; then to 
$\sim$B9 (shown in Fig.~2) coinciding with the peak of 1984 August; and back 
again to $\sim$B7 in 1985 January during the decline from the 1984 peak.  
Then, as the visual light continued to increase, the full A-type spectrum 
developed in 1986 (Wolf et~al.\ 1988) and was maintained through 1998 
(Figs.~2--3).  

These relationships between the spectrum and lightcurve encourage a 
detailed comparison between the spectral types and apparent magnitudes,
which is given in Table~2.  It is seen that the correlation is indeed very 
well defined, and that it applies equally to both the rise and decline, 
insofar as they have been covered.  This correlation is, of course, a
result of the ``constant'' bolometric luminosity during the outburst, as
shown in the table.  The scatter in the derived bolometric magnitudes is not 
surprising, given the different sources of the corrections and the fact that 
they are computed for normal supergiants, as well as the uncertainties in
the classification of these peculiar spectra.  

Even more detailed effects occur, e.g., the apparent doubling of the 
Mg~II line during 1991 (Fig.~2), when the small decline from the 
maximum hump (with an upward spike?) to an extended plateau took 
place (Fig.~5).  It is quite possible that other small differences among 
the spectra are related to spectroscopically unresolved fine structure in 
the lightcurve.  A larger spectral change is evidently related to the 
steep decline in light beginning in 1999 (Figs.~3--4).  There is extensive
spectroscopic coverage during 2002, not shown here, which may elucidate the 
strong spike in light during that year; indeed, the Mg~II line again doubled 
during September, near the lightcurve peak.  The 2002 observation shown 
might again be classified as $\sim$B7, from the He~I/Mg~II absorption ratio, 
albeit with residual A-type emission lines.  

A striking feature of Figure~3 is the apparent monotonic broadening of the 
A-type lines as they fade, which is caused by the gradual dominance of 
[Fe~II] further out in the wind over Fe~II at nearly the same wavelengths, 
as discussed by Stahl et~al.\ (2001) in AG~Car.  In contrast, the He~I 
$\lambda$4471 emission line was very narrow when it first appeared in 2005.  
The recent spectra also display changing structure in the Balmer P~Cyg 
absorption components, not shown here, analogous to what was observed during 
the R127 rise and in AG~Car.  All of these features in R127 will be measured 
and discussed in subsequent publications.  

It is now vital to increase the frequency of observations during the coming 
years, to document R127's final decline to the minimum, quiescent state more 
thoroughly.  In particular, it will be of some interest to catch the 
reappearance of the He~II and N~III Of emission lines--provided that an 
infrared catastrophe as in $\eta$~Car does not obscure the star entirely.
It is possible that the high-ionization lines may already be very weakly
present in the 2008 February and May observations.

\acknowledgments
We thank the Carnegie Supernova Project team, in particular M.~Phillips, 
W.~Krzemin\-ski, Sergio Gonz\'alez, and Carlos Contreras, for the LCO Swope
photometric images.  This paper includes an observation from a 6.5m Magellan 
Telescope at LCO, Chile.  The paper is in large part based on observations 
collected by multiple programs at both the La Silla and Paranal sites of the 
European Organization for Astronomical Research in the Southern Hemisphere, 
Chile.  The Complejo Astron\'omico El Leoncito (CASLEO) in San Juan, Argentina 
is operated under agreement between CONICET, SeCyT, and the Universities of 
La Plata, C\'ordoba, and San Juan.  The CCD and data-acquisition system 
used there are supported under US NSF grant AST-90-15827 to R.M.\ Rich.  
We acknowledge with thanks the variable star observations from the AAVSO 
International Database contributed by observers worldwide and used in this 
research.  Publication support was provided by NASA through grant GO-11212.03-A from STScI, which is operated by AURA, Inc., under NASA contract NAS 5-26555.  Finally, we thank the anonymous referee for requesting that the effects of the companion on the LBV photometry be investigated, which improved the accuracy of the presentation.


\newpage

\begin{deluxetable}{lcccc}
\tablenum{1}
\tablewidth{0pt}
\tablehead{
\colhead{Date} &\colhead{Telescope} &\colhead{Instr} &\colhead{R}
&\colhead{PI/Observer}}
\startdata
1984 Aug 31   &ESO 3.6m     &CASPEC   &20,000   &B.\ Wolf\\
1986 Aug 23    &"            &"        &"         &"\\
1987 Nov  6    &"            &"        &"         &"\\
1989 Jan 22    &"            &"        &"         &OS\\
1989 Dec 16    &"            &"        &"       &B.\ Wolf\\
1991 Jan 28    &"            &"        &"         &"\\
1991 Dec 16    &"            &"        &"         &"\\
1993 Dec 23    &"            &"        &"         &"\\
1995 Feb 16    &"            &"        &"         &"\\
1996 Jan 26    &"            &"        &"         &"\\
1997 Jan 26    &"            &"        &"         &"\\
\\
1998 Oct 11   &ESO 1.5m     &FEROS\tablenotemark{a}    &48,000   &OS\\
1999 Jul 26   &"            &"        &"        &"\\
2002 Sep 18   &ESO 8.0m     &UVES\tablenotemark{b}     &80,000   &ThS\\
2005 Feb 21   &LCO 6.5m     &MIKE\tablenotemark{c}     &60,000   &NS\\
2007 Feb 20   &ESO 2.2m     &FEROS    &48,000   &B.\ Davies\\
2008 Jan 13   &CASLEO 2.2m  &REOSC\tablenotemark{d}    &25,000   &RCG\\
2008 Feb 13   &"            &"        &"        &"\\
2008 May 14   &ESO 2.2m     &FEROS    &48,000   &NIM
\enddata
\tablenotetext{a}{Fiber-fed Extended Range Optical Spectrograph}
\tablenotetext{b}{Ultraviolet-Visual Echelle Spectrograph}
\tablenotetext{c}{Magellan Inamori Kyocera Echelle}
\tablenotetext{d}{Echelle spectrograph on loan from the Institut
d'Astrophysique de Li\`ege}
\end{deluxetable}

\newpage

\begin{deluxetable}{ccclc}
\tablenum{2}
\tablewidth{0pt}
\tablehead{
\colhead{Vis Mag} &\colhead{Sp Type} &\colhead{Epochs}
&\colhead{\phn BC} &\colhead{$m_\mathrm{bol}$\tablenotemark{e}}}
\startdata
11.8--12.2 &Ofpe &$\leq$1978 &$-2.8$\tablenotemark{a} &9.0--9.4\\
10.6--11.0 &B1-2 &1982 Jan, 2007 Feb, 2008 Jan--Feb
&$-1.9$\tablenotemark{b} &8.7--9.1\\
10.2--10.3 &B7 &1983 Feb, 1985 Jan, 2002 Sep? &$-0.6$\tablenotemark{c}
&9.6--9.7\\
9.8 &B9 &1984 Aug &$-0.4$\tablenotemark{c} &9.4\\
8.8--9.5 &A &1986 Mar--1998 Oct &$-0.1$\tablenotemark{d} &8.7--9.4\\
\enddata
\tablenotetext{a}{Martins, Schaerer, \& Hillier 2005 (O9~I)}
\tablenotetext{b}{Lanz \& Hubeny 2007 (20,000 K, lowest g)}
\tablenotetext{c}{Humphreys \& McElroy 1984 (supergiants)}
\tablenotetext{d}{Schiller \& Przybilla 2008 (Deneb, A2~Ia)}
\tablenotetext{e}{With $E_{B-V}=0.20$ and ${V_0}-{M_V}=18.6$,
$m_\mathrm{bol}=9.0$ corresponds to $M_\mathrm{bol}=-10.2$.}
\end{deluxetable}

\newpage

\begin{figure}
\epsscale{1.0}
\plotone{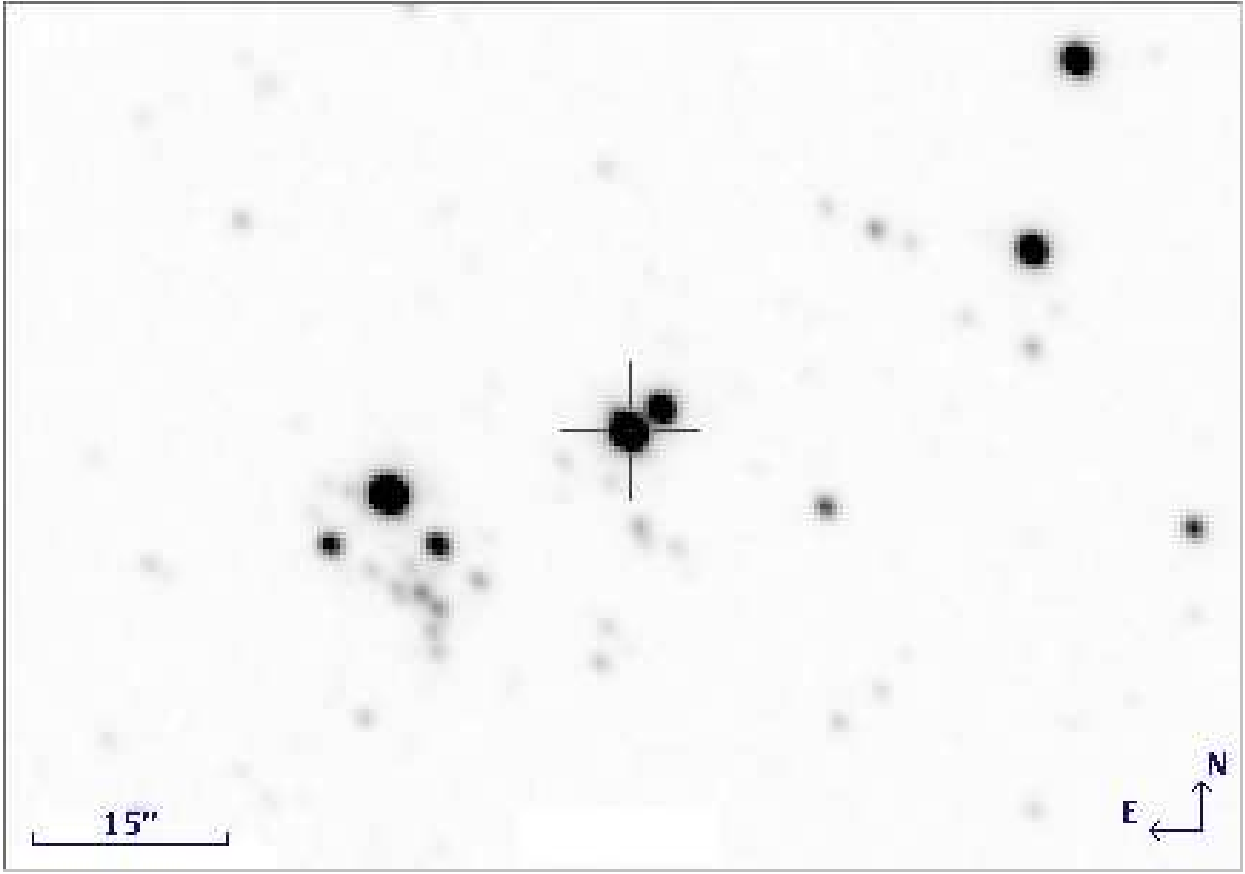}
\caption{\label{fig:fig1}
LCO Swope 1~m, 3~s $V$ exposure of R127 (marked with a cross) and NGC~2055, obtained on 2008 March 19.  The seeing was 1\farcs5.  Compare with the image taken at R127 maximum light, shown by Walborn et~al.\ 1991.  (Note that the scale given in that caption is incorrect.)}
\end{figure}

\begin{figure}
\epsscale{0.7}
\plotone{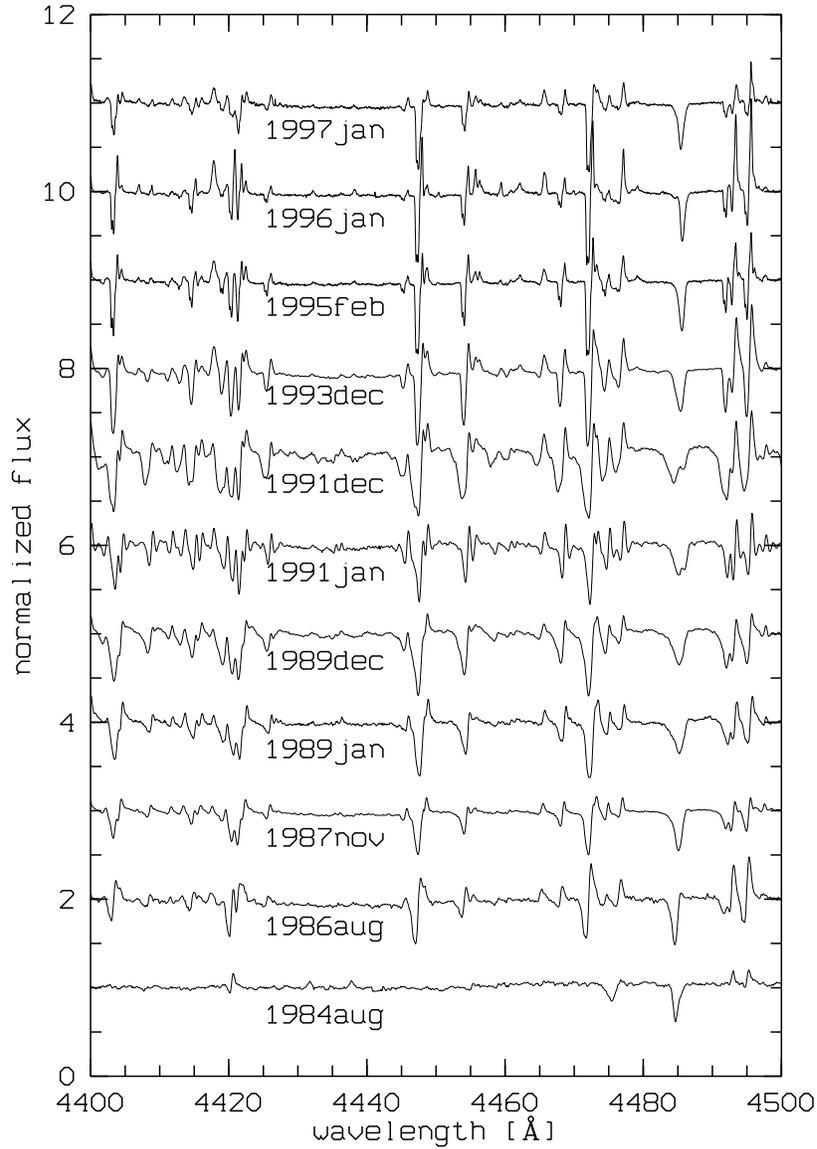}
\caption{\label{fig:fig2}
Time sequence of rectified spectroscopic observations of R127 (see Tab.~1).
The wavelength scale in the present spectroscopic illustrations is
heliocentric, i.e., R127's radial velocity of 293~km~sec$^{-1}$ or
4.4~\AA\ at 4500~\AA\ is included.  The 1984 Aug observation shows strong
He~I $\lambda$4471 and Mg~II $\lambda$4481 absorption lines; the latter
is present throughout and is double in 1991 (see text).  The strongest
P~Cyg profiles are due to Fe~II $\lambda\lambda$4414--17, 4489, 4491 and
Ti~II $\lambda\lambda$4444, 4468.}
\end{figure}

\begin{figure}
\epsscale{0.7}
\plotone{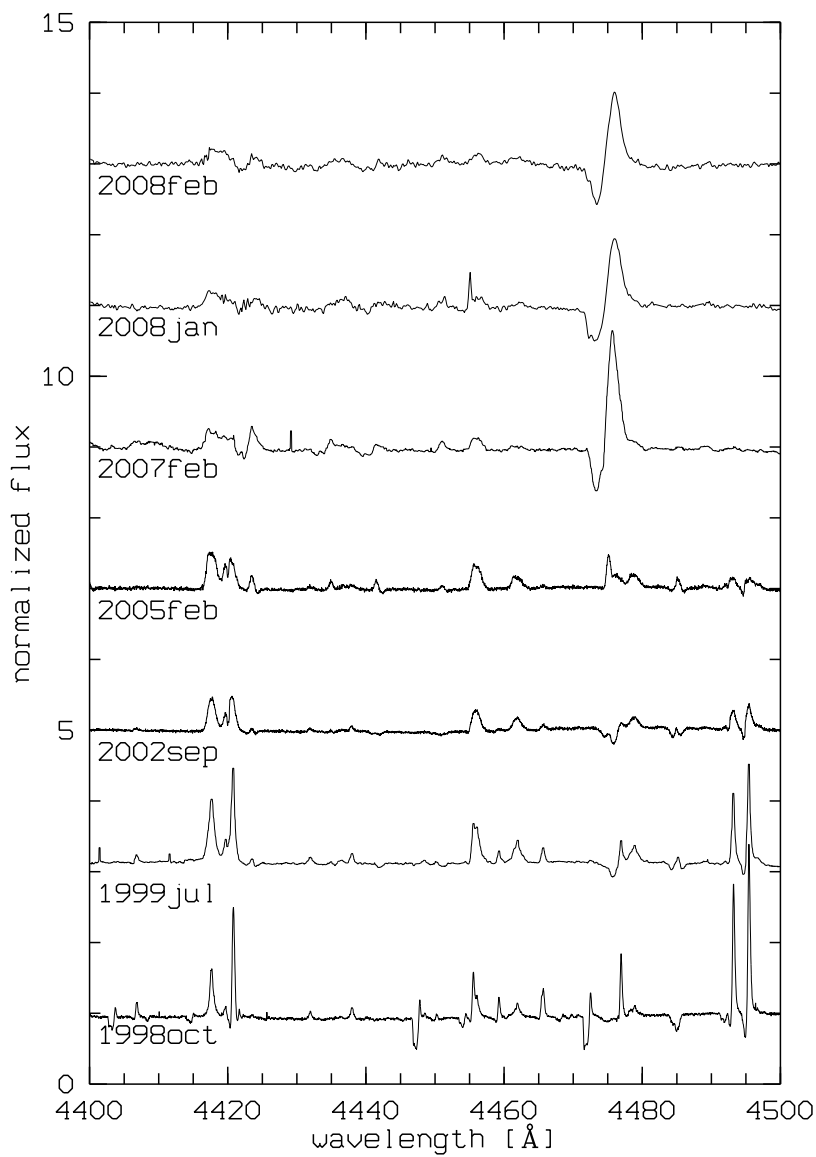}
\caption{\label{fig:fig3}
Continuation of the Fig.~1 time sequence, but with higher resolutions
(Tab.~1).  In 1998 the Ti~II (I.P. 13.6~eV) features weakened substantially 
and they disappeared entirely in 1999, while Fe~II (16.2~eV) remained strong.
The behavior of the 4414--17~\AA\ complex with time is remarkable and is
caused by the increasing contribution of [Fe~II] relative to Fe~II at
nearly the same wavelengths; the former arises further out in the wind
where the velocity is higher.  Note also the transition from Fe~II 
$\lambda$4473 to He~I $\lambda$4471 from 1999 through 2005, and the double 
Mg~II $\lambda$4481 absorption in 2002.  The 2007--2008 spectra are dominated by the very strong He~I $\lambda$4471 P~Cyg profile.}
\end{figure}

\begin{figure}
\epsscale{0.7}
\plotone{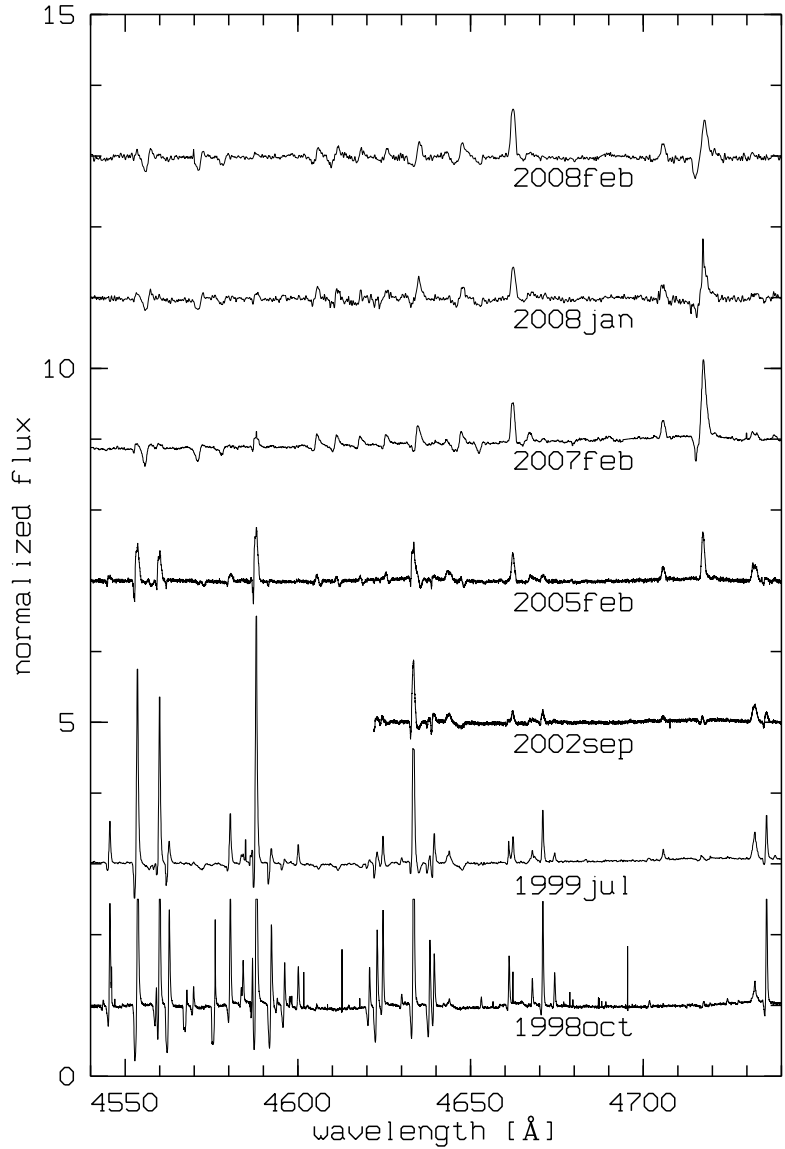}
\caption{\label{fig:fig4}
The same spectrograms as in Fig.~3, at somewhat longer wavelengths.  The 
four strongest emission features in 1998 Oct reach about 6 continuum units 
and have been truncated here, as has the last in 1999 July; they are Fe~II 
$\lambda\lambda$4549, 4556, 4584, 4629.  In addition to the P~Cyg profile of 
He~I $\lambda$4713, the B-type features Si~III $\lambda\lambda$4552, 4568, 
4575 and N~II $\lambda\lambda$4601, 4607, 4614, 4621, 4631, 4643 are prominent in the 2007--2008 spectra, as are [Fe~III] $\lambda\lambda$4658, 4701.5.}
\end{figure}

\begin{figure}
\epsscale{1.0}
\plotone{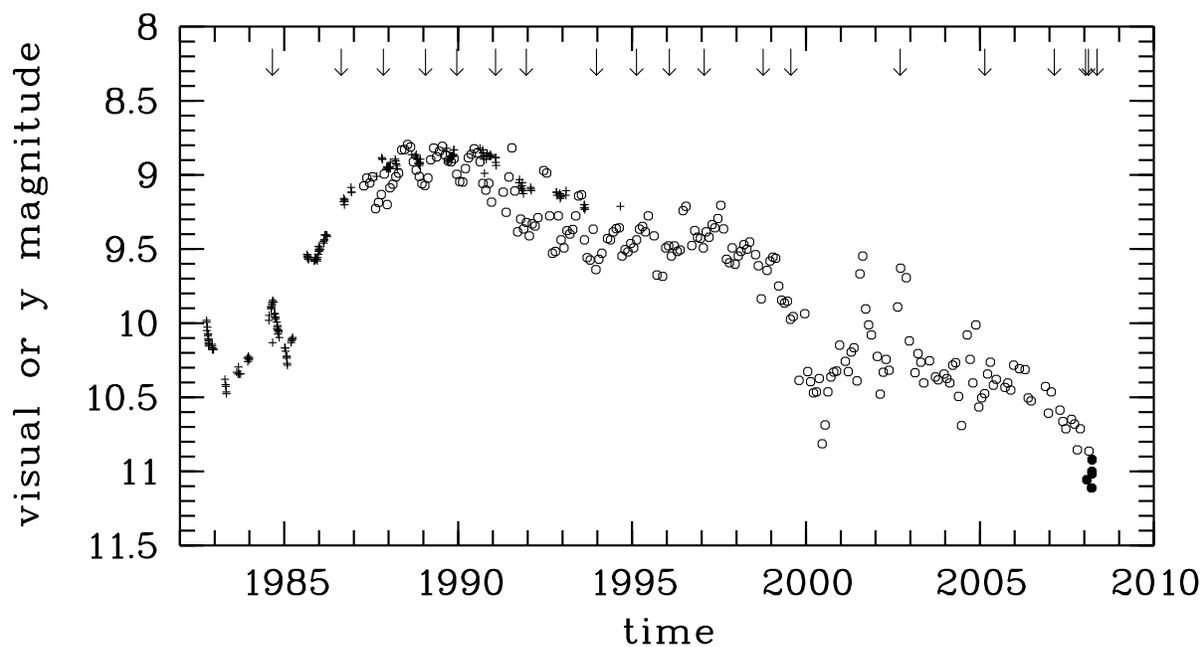}
\caption{\label{fig:fig5}
The composite visual light curve of R127 from 1982 through early 2008.
All measurements have been corrected for the companion.
Crosses: $y$~data from Sterken et~al.\ (1995).  Open circles: monthly
averages of AAVSO visual observations.  Filled squares: new LCO $V$~data.  
The epochs of the spectroscopic observations discussed here are marked by
arrows at the top.}
\end{figure}

\end{document}